# Egonoise Resilient Source Localization and Speech Enhancement for Drones Using a Hybrid Model and Learning-Based Approach

Yihsuan Wu, Yukai Chiu, Michael Anthony, and Mingsian R. Bai, Senior *Member, IEEE*

*Abstract*—Drones are becoming increasingly important in search and rescue missions, and even military operations. While the majority of drones are equipped with camera vision capabilities, the realm of drone audition remains underexplored due to the inherent challenge of mitigating the egonoise generated by the rotors. In this paper, we present a novel technique to address this extremely low signal-to-noise ratio (SNR) problem encountered by the microphone-embedded drones. The technique is implemented using a hybrid approach that combines Array Signal Processing (ASP) and Deep Neural Networks (DNN) to enhance the speech signals captured by a six-microphone uniform circular array mounted on a quadcopter. The system performs localization of the target speaker through beamsteering in conjunction with speech enhancement through a Generalized Sidelobe Canceller-DeepFilterNet 2 (GSC-DF2) system. To validate the system, the DREGON dataset and measured data are employed. Objective evaluations of the proposed hybrid approach demonstrated its superior performance over four baseline methods in the SNR condition as low as -30 dB.

*Index Terms*—Beamsteering, DeepFilterNet, Drone, Generalized Sidelobe Canceller

## I. Introduction

UNMANNED Aerial Vehicles (UAVs), also referred to as drones, are a class of versatile flying machines capable of operating in areas that are difficult to access. The use of drones is increasingly important in search and rescue missions, as well as military operations. These devices have found extensive application in various fields, including agriculture, disaster management, aerial photography, package delivery, and military application, among others. However, the majority of drones rely significantly on camera vision [1]–[4], a limitation that compromises their efficacy in conditions characterized by limited visibility, such as during nighttime or poor weather conditions [5][6].

Although most drones are equipped with camera vision capabilities, drone audio remains under-explored due to the inherent challenge posed by the noise from the rotors. Therefore, the goal of this study is to empower drone control station operators with the audio reality from the First-Person View (FPV). Some research has been dedicated to the fields of speech enhancement [7]–[11] and sound source localization [12]–[16] in the context of drones. This technology can be useful in missions such as search and rescue, particularly during nocturnal operations where cameras may be ineffective. Therefore, it is imperative to mitigate the adverse impacts of drone noise, which is characterized by its high intensity and nonstationary nature. In reality, human speakers can be at a considerable distance from the drone, resulting in extremely low SNRs (–30 dB in many cases).

In order to enhance speech in situations where voice capture is performed over a distance, a speech enhancement system that combines both model-based array signal processing and learning-based neural network approaches is proposed in this paper. The Minimum Variance Distortionless Response (MVDR) beamformer [17], or more generally, the Linearly Constrained Minimum Variance (LCMV) beamformer [18] are two widely used superdirective beamformers. In this paper, we employ the adaptive implementation of MVDR, also known as the Generalized Sidelobe Canceller (GSC) [19], which consists of a fixed beamformer, a blocking matrix, and an adaptive noise canceller, where the Recursive Least Squares (RLS) algorithm [20] is employed in this study. The effectiveness of the beamformers in extracting the source signal while rejecting interference in the non-target directions renders them well-suited for scenarios such as a drone noise scenario [21]. In addition, Schwartz et al. [22] proposed an LCMV beamformer

This work was supported by the National Science and Technology Council (NSTC), Taiwan, under the project number 113-2221-E-007 -057 -MY3. (Corresponding author: Mingsian R. Bai).

Yihsuan Wu was with the Department of Power Mechanical Engineering, National Tsing Hua University, Hsinchu, Taiwan (e-mail: sharon3363451@gmail.com).

Yukai Chiu is with the Department of Power Mechanical Engineering, National Tsing Hua University, Hsinchu, Taiwan (e-mail: kevinchiu500@gmail.com).

Michael Anthony is with the Department of Electrical Engineering, National Tsing Hua University, Hsinchu, Taiwan (e-mail: michaelzhang220@gmail.com).

Mingsian R. Bai is with the Department of Power Mechanical Engineering and Electrical Engineering, National Tsing Hua University, Hsinchu, Taiwan (e-mail: msbai@pme.nthu.edu.tw).

with a Wiener postfilter for multi-speaker separation. Cohen [23] suggested a two-channel GSC with postfiltering for the enhancement of speech corrupted with non-stationary noises.

In addition to the aforementioned ASP-based preprocessing, a learning-based backend is employed to boost performance through a lightweight architecture – the so-called "hybrid approach" [24][25]. In this study, DeepFilterNet 2 (DF2) [26][27] is adopted as a postfilter of the preceding GSC module. DF2 is a low-complexity but extremely effective network [28][29] that was proposed by Schröter et al. for speech enhancement. DF2 utilizes an architecture comprising an encoder and two decoders. One decoder generates a mask in the Equivalent Rectangular Bandwidth (ERB) domain to process the speech envelope according to human auditory perception, while another decoder predicts linear filter coefficients in the low-frequency Short Time Fourier Transform (STFT) domain. While the majority of learning-based enhancement approaches are intended for relatively mild SNR conditions (above −5 dB) [30], Tan et al. [31] attempted to address the very low-SNR problem in the context of drone noise through a compact dilated convolutional neural network (CNN), with a large analysis window to achieve high spectral resolution tailored to the narrow-band harmonic components. Wang and Cavallaro [10] presented a hybrid approach that employs a DNN to estimate a time-frequency mask for speech and noise spatial covariance matrix (SCM) computation, as required by a multichannel wiener filter (MWF). Mukhutdinov et al. [5] compared twelve DNN models for extremely low-SNR scenarios (-30dB) and found that the time-frequency (TF) domain U-Net encoder-decoder architectures provide the best compromise between speech quality, model size, and computational efficiency. DF2 falls into this category. The GSC-DF2 was demonstrated to be effective in enhancing speech signals captured by a six-microphone uniform circular array on a quadcopter. Experimental results demonstrate high-quality speech enhancement performance achievable at an SNR as low as −30 dB.

The paper is organized as follows. The formulation of the problem is presented in Section II. Next, Section III delineates the proposed architecture of the hybrid ASP-DNN system. Section IV details the experimental setup and results. Conclusions and future work are addressed in Section V.

## II. PROBLEM FORMULATION

Consider an $M$-microphone array mounted on a drone. Assuming that the sound emitted from a single source positioned in the far field and captured by the $m$-th microphone can be written in the STFT domain as

$$X_m(l,k) = A_m(l,k)S(l,k) + V_m(l,k) \quad (1)$$

where $m = 1, 2, …, M$, the integers $l$ and $k$ denote the time frame index and the frequency bin index. $S(l, k)$, $X_m(l, k)$, $A_m(l, k)$, and $V_m(l, k)$ denote the clean speech signal, the noisy signal, the acoustic transfer function (ATF), and the rotor noise associated with the $m$-th microphone. It follows that the array signal model can be expressed in the following vector form:

$$\mathbf{x}(l,k) = \mathbf{a}(l,k)S(l,k) + \mathbf{v}(l,k) \quad (2)$$

where $\mathbf{a}(l, k) = [A_1(l, k)\ A_2(l, k)\ …\ A_M(l, k)] = [e^{-j\mathbf{k}\cdot\mathbf{r}_1}\ e^{-j\mathbf{k}\cdot\mathbf{r}_2}\ …\ e^{-j\mathbf{k}\cdot\mathbf{r}_M}]^T$ being the steering vector based on the freefield plane-wave ATF model, as drones are usually operated in outdoor open space. The superscript "$T$" denotes matrix transposition. The wave vector $\mathbf{k} = -(\omega/c)\boldsymbol{\kappa}$, with $\omega$ being the angular frequency, $c$ being the speed of sound and $\boldsymbol{\kappa}$ being the unit vector pointing at the look direction. The noise vector, $\mathbf{v}(l, k) = [V_1(l, k)\ V_2(l, k)\ …\ V_M(l, k)]^T$, represents the egonoise vector of the rotors. The objective of this study is to recover the speech signal $S(l, k)$ that has been corrupted by the significantly stronger rotor noise $\mathbf{v}(l, k)$ from the noisy microphone signals $\mathbf{x}(l, k)$.

## III. PROPOSED METHOD

In this work, the very low-SNR issue caused by drone rotor noise is addressed through the implementation of a hybrid ASP-DNN approach, as illustrated in Fig. 1. The proposed system is comprised of a model-based GSC front end, followed by a learning-based postfilter. For simplicity, the TF argument (l, k) is omitted hereafter.

The GSC is comprised of two branches. The upper branch extracts the target signal using a fixed beamformer, while the lower branch utilizes a blocking matrix and an adaptive noise canceller to suppress interference (rotor noise) leaking from the non-target directions. That is,

$$\mathbf{w}_{GSC} = \mathbf{w}_c - \mathbf{B}\mathbf{w}_a \quad (3)$$

where $\mathbf{w}_c$ denotes the weight vector of the fixed beamformer, $\mathbf{B}$ denotes the blocking matrix that blocks the target signal, and $\mathbf{w}_a$ denotes the coefficient vector of the adaptive noise canceller. As previously stated, the freefield plane-wave ATF model is employed to construct $\mathbf{w}_c$ and $\mathbf{B}$. This feature facilitates beamsteering in direction of arrival (DOA) estimation. The GSC output serves as an input feature to a learning-based postfilter, as will be detailed next.

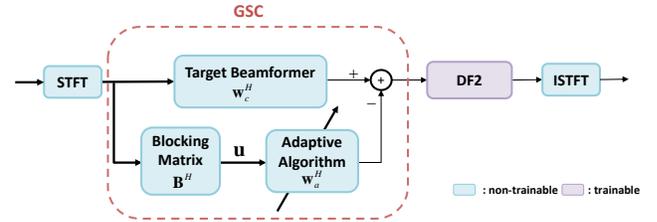

Fig. 1. The proposed microphone array signal processing architecture for very low SNR speech enhancement.

### A. The GSC front end

Consider a single source setting. We implement a GSC beamformer to extract the target source signal with a Delay-and-Sum Beamformer (DSB) in its upper branch, as shown in Fig. 1. Under the freefield condition, the weight vector of the fixed beamformer is given by

$$\mathbf{w}_c = \frac{\mathbf{a}(\theta_i)}{\mathbf{a}^H(\theta_i)\mathbf{a}(\theta_i)} = \frac{\mathbf{a}(\theta_i)}{M} \quad (4)$$

The blocking matrix in the lower branch of the GSC is formulated using the projection method. The superscript "$H$" denotes matrix transpose conjugation. The projection matrix is defined as

$$\mathbf{B} = \mathbf{I} - \frac{\mathbf{a}(\theta_i)\mathbf{a}^H(\theta_i)}{\mathbf{a}^H(\theta_i)\mathbf{a}(\theta_i)} \quad (5)$$

The blocking matrix $\mathbf{B}$ projects the microphone signals onto the subspace that is orthogonal to target source direction, or the subspace that supposedly accounts for the rotor noise, $\mathbf{I}$ is the identity matrix, $\mathbf{a}(\theta_i)$ is the steering vector in the target direction $\theta_i$.

Unlike conventional GSC, which uses the Normalized Least-Mean-Squares algorithm for adaptive filtering, Recursive Least Squares (RLS) is employed due to its faster convergence property. This property is especially vital for non-stationary rotor noise. Next, the blocked microphone signal $\mathbf{u}(l)$ is filtered by the adaptive filter $\mathbf{w}_a(l)$ to yield the noise prediction $\hat{d}(l)$. Thus, the *a posteriori* error signal, or equivalently a GSC-enhanced signals, can be written as

$$e(l) = d(l) - \hat{d}(l) = d(l) - \hat{\mathbf{w}}_a^H(l)\mathbf{u}(l) \quad (6)$$

The weight update procedure for the RLS algorithm is summarized as follows [39]:

$$\xi(l) = d(l) - \hat{\mathbf{w}}_a^H(l-1)\mathbf{u}(l) \quad (7)$$

$$\mathbf{k}(l) = \frac{\lambda^{-1}\mathbf{P}(l-1)\mathbf{u}(l)}{1+\lambda^{-1}\mathbf{u}^H(l)\mathbf{P}(l-1)\mathbf{u}(l)} \quad (8)$$

$$\hat{\mathbf{w}}_a^H(l) = \hat{\mathbf{w}}_a^H(l-1) + \mathbf{k}(l)\xi^*(l) \quad (9)$$

$$\mathbf{P}(l) = \lambda^{-1}\mathbf{P}(l-1) - \lambda^{-1}\mathbf{k}(l)\mathbf{u}^H(l)\mathbf{P}(l-1) \quad (10)$$

where $\xi(l)$ is the *a priori* estimation error, the $M$-by-1 vector $\mathbf{k}(l)$ is the Kalman gain vector, the $M$-by-$M$ matrix $\mathbf{P}(l)$ is the inverse covariance matrix, $\lambda$ is the forgetting factor, which is typically chosen in the range of 0.98 to 1. There is one caveat to the matrix $\mathbf{P}(l)$ which is defined as the inverse of the covariance matrix of the blocked signal in the RLS recursion above. To ensure numerical stability in computing $\mathbf{P}(l)$, it is necessary to delete the last column of the blocking matrix in Eq. (5) prior to performing the RLS recursion.

*B. The DF2 back end*

The DF2 [26] is a low-complexity speech enhancement model that is designed for real-time processing on embedded devices. The utilization of such a network offers distinct advantages for applications such as drones, where real-time audio processing is imperative and hardware limitations necessitate low-complexity solutions. DF2 employs an encoder-decoder architecture. Two decoders are in operation: one in the ERB domain and the other in the STFT domain. The first decoder generates a real-valued magnitude mask in 32 ERB bands. The second decoder generates linear filter coefficients in the STFT domain for low frequencies, where speech periodicity, linked to tone and pitch, is most prominent. These deep filtering coefficients are applied to the previously stated masked signals to focus on energy-dense regions of speech and improve clarity. It has been demonstrated that this filtration strategy exhibits superior performance in low-SNR scenarios when compared to conventional complex ratio masks.

In this study, we build upon the pretrained DF2 model [26] and adapt it to address the present speech enhancement in microphone array embedded drone scenario. Specifically, the pretrained network model is "fine-tuned" using the GSC outputs derived from recorded drone noise and the open-source DREGON dataset [35]. This refinement enables the model to become more resilient to the spectral characteristics and variability of real-world drone noise, thereby enhancing its generalizability to unseen flight scenarios even under extremely low-SNR conditions encountered in UAV applications. The Adam optimizer is employed for model training. All other training settings adhere to the initial configuration outlined in [26]. This model entails 2.31 million parameters, a computational complexity of 0.36 G MACs per second, and a real-time factor (RTF) of 0.04.

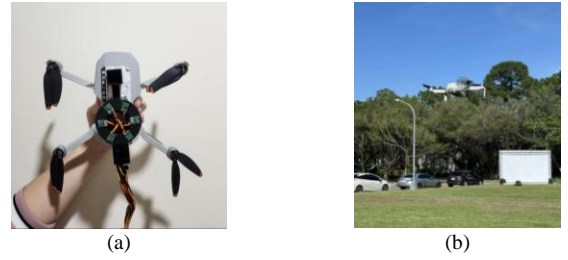

Fig. 2. Experiment setup. (a) Top view of the drone with a circular microphone array, (b) Drone-based outdoor data acquisition scenario.

GSC and DF2 in the suggested hybrid approach are based on existing systems, yet the following study will show that combining these two elements properly results in much improved performance compared to previous methods.

## IV. EXPERIMENTAL SETUP AND RESULTS

In order to validate the proposed hybrid drone localization and enhancement system, experiments were performed and the results were compared to those of four baseline methods.

*A. Experimental Settings*

The target speech employed in the experiment is derived from the "train-clean-100" subset of the LibriSpeech corpus [34], which contains 100 hours of clean speech from male and female speakers, sampled at 16 kHz. The drone noise was recorded using a circular array of 3.5 cm radius with six analog Micro-Electro-Mechanical Systems (MEMS) microphones uniformly distributed on a Plexiglas plate (Fig. 2). The microphone array was mounted on a quadcopter, DJI® Mini 2. The audio data was recorded using a PreSonus® audio interface.

The speech signals are pre-mixed with recorded drone noise on randomly selected SNR levels ranging from −5 dB to −30 dB in 5 dB steps. In the training stage, a total of 30,000 noisy signal clips were generated for training. An additional 3,000 samples, which the models had not seen, were also generated for validation. This was done to assess the generalizability of the model. In the testing stage, a total of 2,000 noisy samples were randomly selected from drone noise data in the DREGON dataset [35] and our recorded data. Each drone noise clip is 4 seconds in duration, with a 2-second segment of speech signal randomly inserted within the clip.

*B. Results and Discussion*

Objective metrics including Perceptual Evaluation of Speech Quality (PESQ), Short-Time Objective Intelligibility (STOI), Scale-Invariant Signal-to-Distortion Ratio (SI-SDR), and SNR are employed for a comprehensive evaluation. The proposed approach is benchmarked against four established baselines. The first baseline is a Dual-stage Multichannel Wiener Filtering (DMWF) [36], which applies MWF to suppress drone noise and then a Gaussian Mixture Model-based Wiener Filter (GMM-WF) for further speech enhancement [37]. The second baseline is a time-frequency masked (TFM) MWF approach based on [7], which estimates the DOA using spatial likelihood functions for each TF bin and applies Gaussian weights based on DOA proximity within an MWF to suppress egonoise while extracting the target signal. We also include a simple GSC and an end-to-end DF2 as two additional baselines.

The localization and enhancement algorithms were evaluated at SNR levels of −10, −20, and −30 dB. Due to space limitations, Figs. 3(a), (b) illustrates only the spectrograms of the clean signal and the noisy signal at an SNR level of −30 dB and a target source located at 180°. Fig. 3(c)(d) shows the localization result and the enhanced result for the lowest SNR condition (−30 dB). The findings demonstrate the superior localization accuracy and enhancement quality achieved using the proposed method at the extremely low SNR condition of rotor noise. A A comprehensive comparison of the performance difference, denoted with a lowercase "d", across objective metrics (dPESQ, dSTOI, dSI-SDR, and dSNR) for input SNR levels ranging −30 to −5 dB with 5 dB steps is illustrated in Fig. 4. The proposed GSC-DF2 consistently outperformed the baselines in all evaluated SNR conditions. In particular, the dSNR reached 72 dB using the proposed method under −30 dB SNR, which is indeed a significant advance over the baselines.

The second best method varies depending on the evaluated metrics. The MWF-GMM and TFM-MWF methods were originally designed for moderate drone noise conditions (around −20 dB and −15 dB, respectively). However, their performance considerably degrades above −20 dB. The first one needs voice activity detection (VAD), while the second one needs a precise estimate of the DOA. Although DF2 has been shown to be effective in high-SNR settings, a notable degradation in enhancement performance is evident in low-SNR scenarios. Lastly, while GSC is a lightweight, free of training, low-complexity, model-based approach, its effectiveness in suppressing noise appears to be insufficient.

The preceding results demonstrate the efficacy of the proposed method, even under conditions of SNR = −30 dB. This prompts the following question: What is the effective detection distance that can be achieved by the proposed method? To answer this question, note that

$$L_{source}(1\ \text{m}) - 20\log_{10} r - L_{drone} \geq SNR_{th,1mic} \text{(dB)} \quad (11)$$

where $L_{source}(1\ \text{m})$ represents the sound pressure level (SPL) of the source at 1 m, $L_{drone}$ represents the SPL of the drone's egonoise picked up at the microphone, $r$ is the distance between the source and the array on the drone, $SNR_{th,1mic}$ denotes the input SNR threshold, and the subscript "th" refers to the threshold at which beamforming ceases to function properly. Thus, it is straightforward to show that, when the equality holds, the effective detection distance is

$$r_{eff} = 10^\delta, \text{ with } \delta = \left[L_{source}(1\ \text{m}) - L_{drone} - SNR_{th,1mic}\right]/20 \quad (12)$$

In this experiment, the effective detection distance as predicted by Eq. (11) is 100 m if $L_{source}(1\ \text{m}) = 90\text{dB}$, $L_{drone} = 80\text{dB}$, $SNR_{th,1mic} = -30\text{dB}$, which amounts to 72 dB dSNR.

## V. CONCLUSIONS

This paper proposes an egonoise-resilient hybrid system for drones. As demonstrated by the results, the system is capable of effective localization and enhancement in an extremely low SNR of −30 dB. This represents a substantial improvement over four prior techniques in the literature. The paper integrates a model-based GSC frontend and a learning-based DF2 backend. The experimental results have confirmed that the proposed system is capable of substantial speech enhancement when assessed with four objective performance metrics in very low SNR conditions due to rotor noise of UAVs. In the future, we intend to expand the current system to accommodate multiple sources with audiovisual localization and enhancement, as well as binaural rendering.

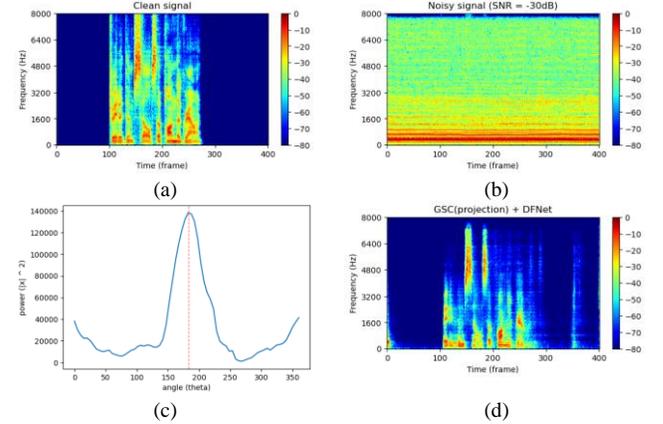

Fig. 3. Results at an SNR level of −30 dB: (a) Clean speech signal, (b) Noisy signal, (c) Enhancement result – power versus angle plot, (d) Enhancement result – Spectrogram of enhancement result using GSC with DF2 post-filtering.

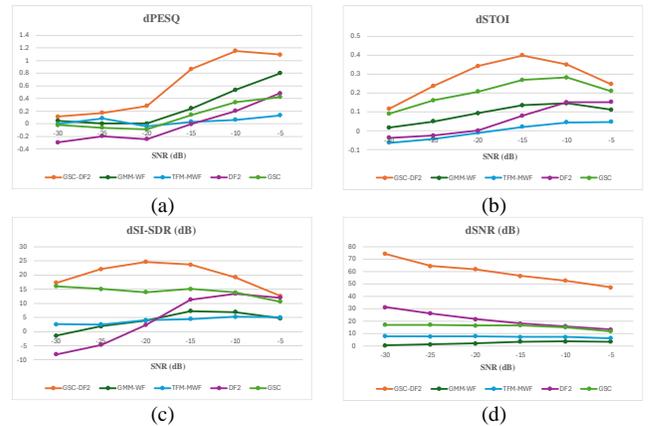

Fig. 4. Comparisons of the proposed method versus baselines under different SNR levels (a) dPESQ, (b) dSTOI, (c) dSI-SDR, (d) dSNR using GSC with DF2 post-filtering.